\documentclass[aps,prl,twocolumn,epsfig] {revtex4}
\usepackage{pdfsync}
\usepackage{graphicx}   
\usepackage{epstopdf}
\usepackage{amsmath}

\begin{document}

\title{The Biaxial Smectic-$A^*$ Phase -- A New Phase, Already But Unknowingly Discovered?}

\author{Karl Saunders}

\affiliation{Department of Physics, California Polytechnic State University, San Luis Obispo, CA 93407, USA}




\begin{abstract}

The biaxial smectic-$A^*$ (Sm-$A_B^*$) phase, appearing in the phase sequence Sm-$A^*$--Sm-$A^*_B$--Sm-$C^*$, is analyzed using Landau theory. It is found to possess a helical superstructure with a pitch that is significantly shorter than the pitch of the Sm-$C^*$ helical superstructure. The Sm-$A_B^*$--Sm-$C^*$ transition can be either 1st or 2nd order, and correspondingly there will be either a jump or continuous variation in the pitch. The behaviors of the birefringence and electroclinic effect are analyzed and found to be similar to those of a Sm-$C^*_\alpha$ phase . As such, it is possible that the Sm-$A_B^*$ phase could be misidentified as a Sm-$C^*_\alpha$ phase. Ways to distinguish the two phases are discussed. 

\end{abstract}


\maketitle

\tighten

\begin{figure}
\includegraphics[scale=0.3]{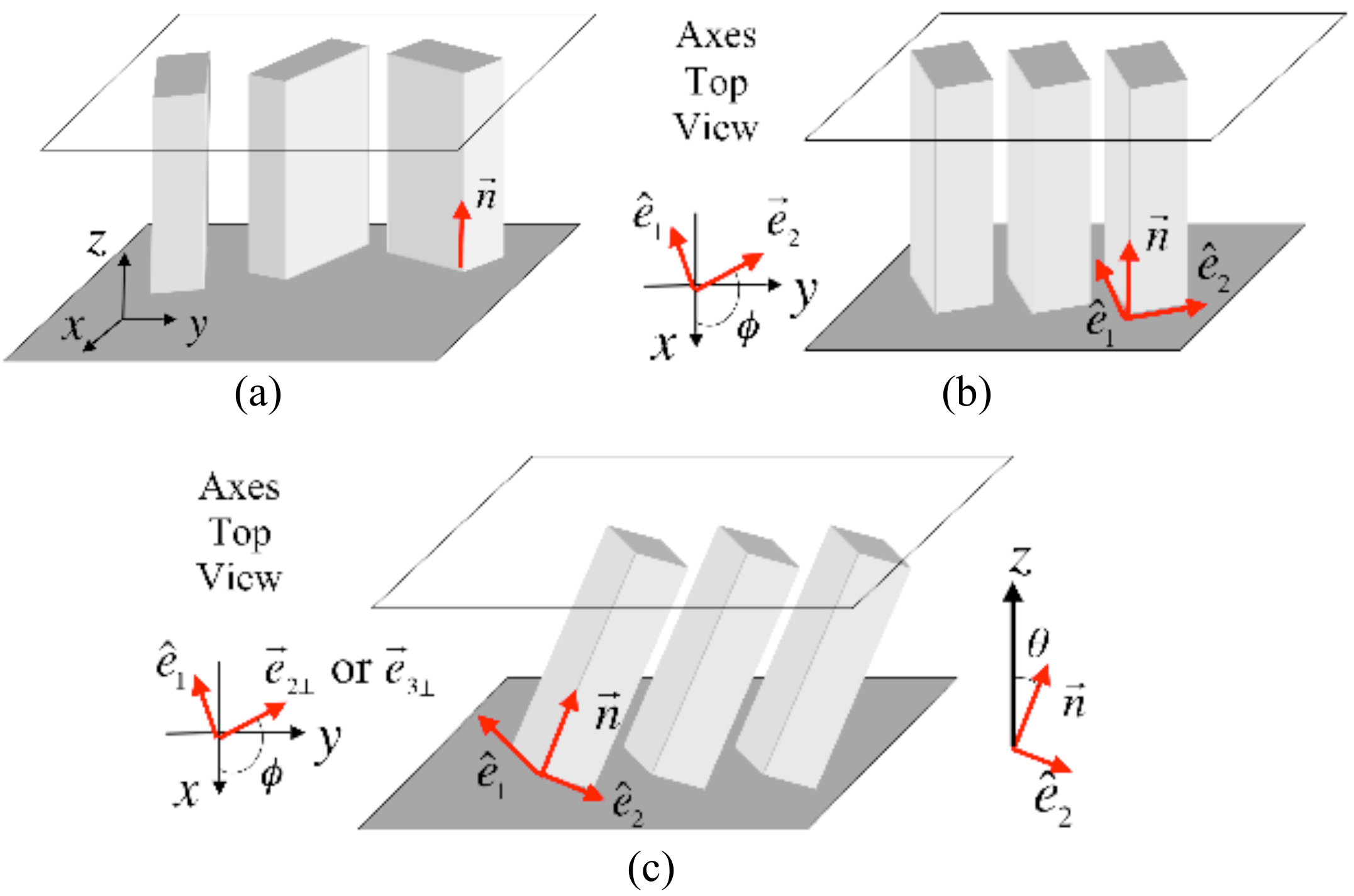}
\caption{Schematics of  (a) Sm-$A^*$ (b) Sm-$A_B^*$ (c) Sm-$C^*$ phases. In each case a single layer is shown. ${\bf \hat e_1}$, ${\bf \hat e_2}$, ${\bf \hat n}$ are the eigenvectors of the orientational order tensor.
\vspace{-30 pt}
}
\label{Schematic}
\end{figure}

Liquid crystals are a fascinating class of materials exhibiting a range of phases (intermediate between liquid and crystalline) which can be classified according to their broken symmetries. The rich variety of their order parameters and phase transitions has led to considerable interest in their properties \cite{ppm}. In condensed matter physics they provide an opportunity to study fundamental issues such as the interplay of different types of order, and the effects of chirality on phases and phase transitions, particularly amongst chiral smectic (Sm$^*$) phases. There is a rich variety of such phases, which are typically made up of elongated molecules, and have a density periodic in one dimension ($\bf \hat z$), i.e. layering \cite{deGennes_book}. As shown in Fig.~\ref{Schematic}, Sm-$A^*$ phases have an average molecular long axis ($\bf \hat n$) parallel to the layer normal ($\bf \hat z$). In lower temperature Sm-$C^*$ phases $\bf \hat n$ is tilted by an angle $\theta$ from $\bf \hat z$. This tilt can be induced by an electric field, a chiral phenomenon known as the electroclinic effect (EE) \cite{EE,BH, Saunders2}. The EE allows for rapid switching of the optical axis ($\bf \hat n$), an important feature for electrooptical devices. The chirality of the Sm-$C^*$ phases results in a helical precession (along $\bf \hat z$) of $\bf \hat n$, with pitch $p_C$ larger than the layer spacing. Thus, as well as layering, Sm-$C^*$ phases have a helical, superstructure which can be probed by Bragg scattering. 

The discovery \cite{EE} of the EE has led to the ongoing synthesis of an enormous number of chiral liquid crystal compounds with smectic phases, also known as ferroelectric liquid crystals. A large fraction of these compounds display a variety of short pitch Sm-$C^*$ phases (as well as the conventional, longer pitch Sm-$C^*$ phase). The ``ferrielectric" (ferri) phases (e.g. Sm-$C^*_{FI1}$ and Sm-$C^*_{FI2}$, with pitches of 3 and 4 layers, respectively) are believed to result from a competition between  ferro- and antiferroelectric interactions \cite{Review}. As such, they are analogous to ferrimagnetic phases, and have been modeled with competing nearest and next-nearest layer interactions \cite{Review}, an example of how a single, fundamental aspect of physics can result in a similar effect in two ostensibly very different systems (magnetic and liquid crystalline). There has also been significant interest \cite{Review} in the Sm-$C^*_\alpha$ ferroelectric phase, which has a pitch between that of the ferri and conventional Sm-$C^*$ phases. However, unlike the ferri phases, its pitch is incommensurate with the layer spacing. It and the ferri phases appear in the phase sequence Sm-$A^*$--Sm-$C^*_\alpha$--Sm-$C^*$--Sm-$C^*_{FI2}$--Sm-$C^*_{FI1}$, with the Sm-$A^*$ phase at highest temperature. The short pitch nature of the Sm-$C^*_\alpha$ phase would naturally lead one to first suspect (as many have \cite{Review}) that, like the ferri phases, it is simply another phase with competing interactions. 

In this Letter we present the first analysis of the chiral biaxial smectic-$A^*$ (Sm-$A^*_B$) phase \cite{Pikin, Osipov Biaxiality}. The Sm-$A^*_B$ and Sm-$C^*_\alpha$ phases have common features: a short pitch helical superstructure, a strong EE effect above the transition to the Sm-$C^*$ phase and also a strong decrease in birefringence below the transition from the Sm-$A^*$ phase. Thus, we suggest that in {\it some} cases a short pitch phase, appearing between the Sm-$A^*$ and Sm-$C^*$ phases, could be mistaken for a Sm-$C^*_\alpha$ phase when it is really a Sm-$A^*_B$ phase. We show that the unusually short pitch and the strong EE of the Sm-$A^*_B$ phase are not due to competing interactions but are due instead to completely different basic physics, namely the distinct symmetries ($D_{2h}$ and $C_{2h}$) of the Sm-$A^*_B$ and Sm-$C^*$ phases. Aside from its obvious scientific and technological importance in terms of better understanding of ferroelectric liquid crystals, this result has a broader significance in terms of the subtleties of phase transitions and phase identification. It demonstrates that two very similar phases can occur for two fundamentally different reasons, competing interactions (Sm-$C^*_\alpha$) and symmetry breaking (Sm-$A^*_B$). In such cases one must be careful to devise methods of distinguishing between two ostensibly similar phases, and we indeed provide such methods. 

That, in {\it some} cases, the supposedly observed Sm-$C^*_\alpha$ phase may really be the Sm-$A^*_B$ phase is also important given that two features of the Sm-$C^*_\alpha$ are puzzling from the point of view of general condensed matter physics. In some materials \cite{Skarabot, Panov} the Sm-$C^*_\alpha$--Sm-$C^*$ phase transition has been observed to be continuous. This would contradict the basic tenet that transitions between phases of the same symmetry must be 1st order \cite{critical point}. Another puzzling feature of the Sm-$C^*_\alpha$ phase is its location in the above phase sequence. One would reasonably expect that the phase sequence of symmetrically equivalent phases should occur in order of decreasing pitch and, therefore, that the Sm-$C^*_\alpha$ phase should appear between the Sm-$C^*$ and the Sm-$C^*_{FI2}$ phases. The existence of a Sm-$A^*_B$ phase could resolve these issues. It and the Sm-$C^*$ phases are symmetrically distinct and a continuous phase transition between the two is permitted. Also, its location in the phase sequence is consistent with it having symmetry between that of the Sm-$A^*$ and Sm-$C^*$ phases.
\begin{figure}
\includegraphics[scale=0.4]{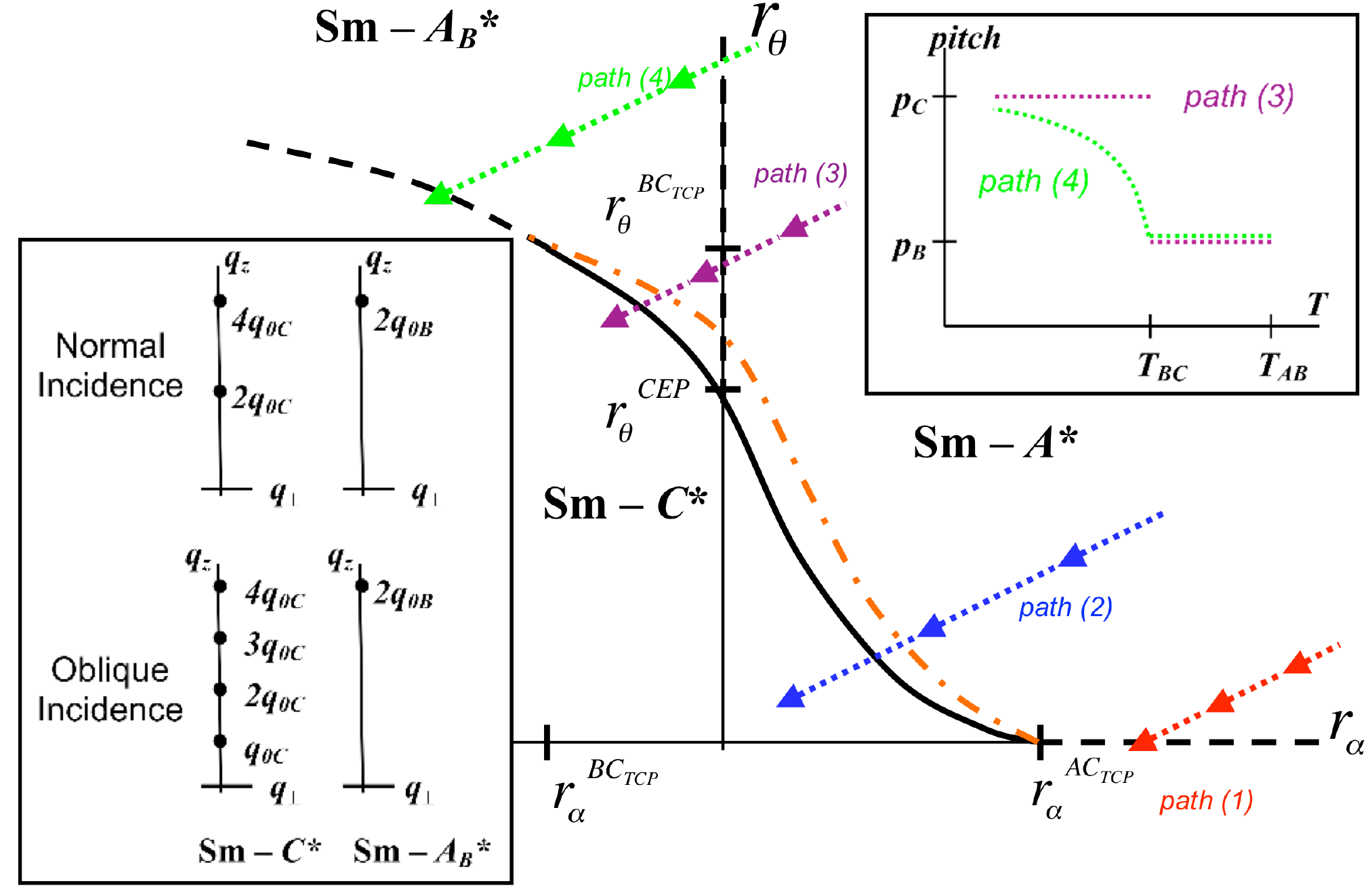}
\caption{ The phase diagram in $r_\alpha$-$r_\theta$ space for the Sm-$A^*$, Sm-$A^*_B$, Sm-$C^*$ phases. 1st and 2nd order phase boundaries are shown as solid and dashed lines respectively. Four decreasing temperature paths from the Sm-$A^*$ to Sm-$C^*$ phase are shown. In the region between the (black) solid and (orange) dot-dashed lines the system will exhibit a particularly particularly dramatic electroclinic effect (see Fig.~\ref{EE}). Upper inset: the expected behavior of the helical pitch across the 1st and 2nd order Sm-$A^*_B$--Sm-$C^*$ transitions. Lower inset: schematic location of the normal or oblique incidence Bragg scattering wavevector peaks associated with the helical superstructure of the Sm-$A^*_B$ or Sm-$C^*$ phases, with $K_b/K_t\approx 2$.
\vspace{-40 pt}
}
\label{PhaseDiagram}
\end{figure}

We first discuss the key features of the Sm-$A^*_B$ phase, along with ways to distinguish it from the Sm-$C^*_\alpha$ phase. As shown in Fig.~\ref{Schematic}, the Sm-$A_B^*$ phase is non-tilted (i.e $\bf \hat n \parallel \bf \hat z$) with a special axis picked out {\it within} the layers. This {\it axis} is usually specified by a biaxial director ${\bf \hat e_1}$ but we note that the Sm-$A_B^*$ phase possesses intra-layer inversion symmetry (i.e. ${\bf \hat e_1}$ and $-{\bf \hat e_1}$ equivalence). In Sm-$C^*$ phases the tilted $\bf \hat n$ picks out a special {\it direction} ${\bf c}$ =${\bf \hat n}-({\bf \hat n}\cdot{\bf \hat z}){\bf \hat z}$ within the layers, and does not possess intra-layer inversion symmetry. Thus, the Sm-$A_B^*$ phase has symmetry between the Sm-$A^*$ and Sm-$C^*$ phases. The chirality of the Sm-$A^*_B$ phase means that the biaxial director ${\bf \hat e_1}$ helically precesses along $\bf \hat z$ with pitch $p_B$. This precession may seem similar to the that of the Sm-$C^*$ phase in which $\bf c$ precesses with pitch $p_C$. However, it will be shown to involve a fundamentally different  helical distortion (twist) than that of Sm-$C^*$ phase (bend). Twist is a lower energy distortion than bend, which explains why the Sm-$A_B^*$ pitch is  shorter than the lower temperature Sm-$C^*$ phase. We show that $p_B$ is up to a factor of $K_b/K_t$ shorter than $p_C$, where $K_b$ and $K_t$ are the nematic twist and bend elastic modulii. Since $K_b/K_t$ is typically of order 2 (and is often more), the Sm-$A^*_B$ pitch will be considerably smaller than the Sm-$C^*$ pitch. 

The Bragg reflections associated with the helical superstructure of the Sm-$A^*_B$ and Sm-$C^*$ (or Sm-$C^*_\alpha$) phases can be distinguished by comparing normal incidence (along $\bf \hat z$) and oblique incidence scattering. Due to the intra-layer inversion symmetry the actual periodicity of the orientational order and associated optical properties of the Sm-$A^*_B$ phase will be $p_B/2$. This is unlike the Sm-$C^*$ phase, which lacks this inversion symmetry and is periodic only over the full pitch $p_C$. It is known \cite{deGennes_book} that the $p_C$ periodicity of the Sm-$C^*$ phase is only revealed for scattering at oblique incidence. As shown in Fig.~\ref{PhaseDiagram}, for normal incidence only Bragg reflections at wavevectors $2nq_0$ (with $q_0 =2 \pi/p_{C/B}$ and $n$ an integer) are observed. Thus, in going from normal to oblique incidence, extra Bragg reflections at odd multiples of $q_0$ will be observed in the Sm-$C^*$ phase but not in the Sm-$A^*_B$ phase. Correspondingly, if measurements are {\it only} made for oblique incidence one may mistake the Sm-$A^*_B$ phase for a Sm-$C^*$ phase with pitch $p_B/2=(K_t/2K_b)p_C$ that is significantly (by a factor of 4 or more) smaller than the actual Sm-$C^*$ phase that appears at lower temperature. 

Another feature of the Sm-$A^*_B$ phase is that its helical superstructure results in a decrease in the birefringence $\Delta n$ from its Sm-$A^*$ value. A similar feature has been observed at the Sm-$A^*$--Sm-$C^*$  and Sm-$A^*$--Sm-$C_\alpha^*$ transitions, and used to obtain $\theta(T)$ via measurements of $\Delta n(T)$ \cite{Skarabot}. If a Sm-$C^*_\alpha$ phase was really a Sm-$A^*_B$ phase then the indirect measurement of the Sm-$C^*_\alpha$ $\theta(T)$, may really be a measurement of the Sm-$A^*_B$ biaxialty $\alpha(T)$.
\begin{figure}
\includegraphics[scale=0.5]{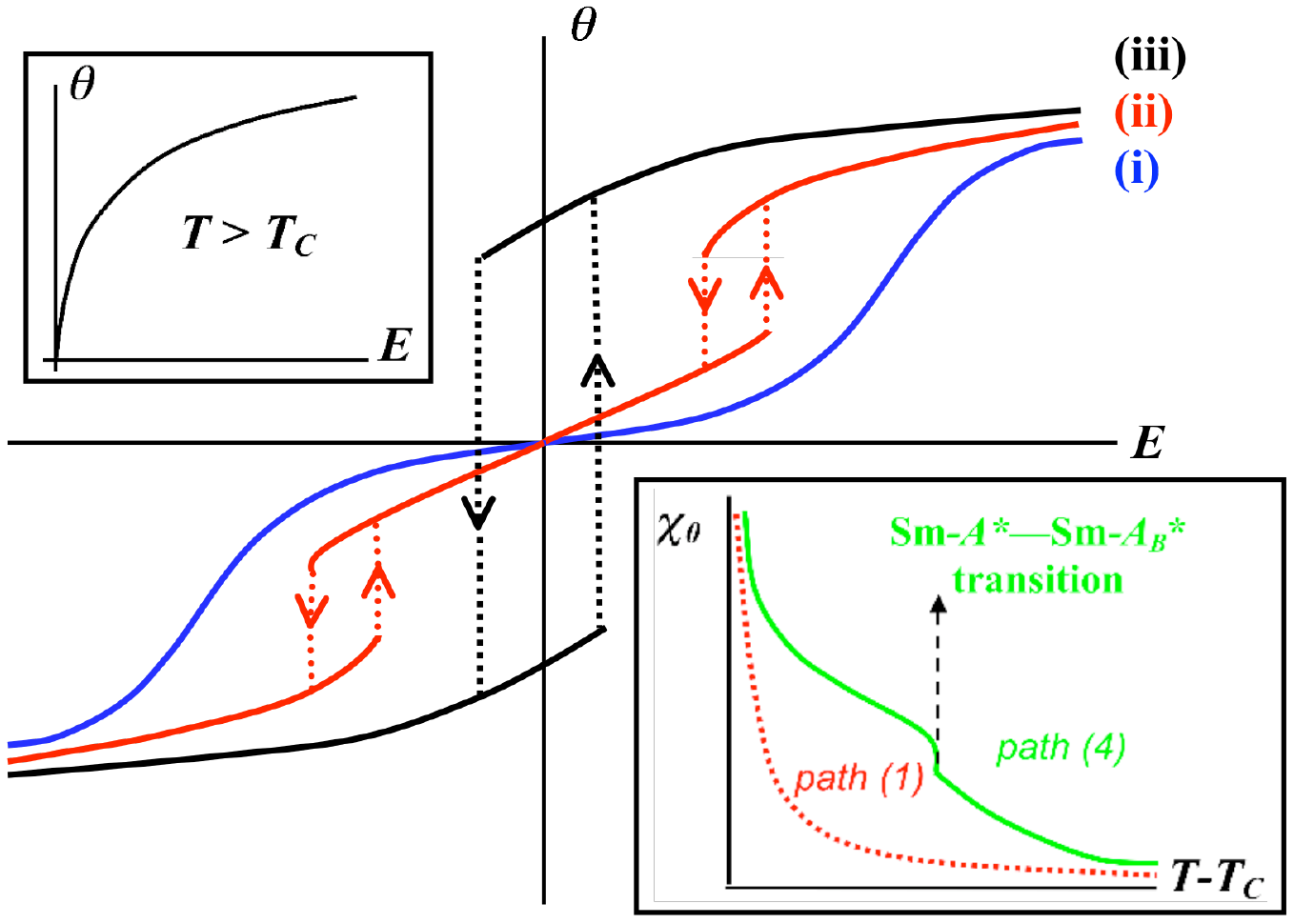}
\caption{Upper Inset: $\theta(E)$ in Sm-$A^*$ or Sm-$A_B^*$ phases above a continuous transition to the Sm-$C^*$ phase. The susceptibility $\chi_0$ is the slope of the $\theta(E)$ curve at $E=0$. Lower Inset: $\chi_0(T)$ for phase sequences Sm-$A^*$--Sm-$C^*$ (red dotted) or Sm-$A^*$--Sm-$A_B^*$--Sm-$C^*$ (green solid). Paths (1) and (4) refer to Fig.~\ref{PhaseDiagram}. The continuous transition to the Sm-$C^*$ phase is at $T_C$. For a 1st order transition the divergence of $\chi_0(T)$ is cut off at $T>T_C$. Main: $\theta(E)$ curves (i) and (ii) are in the Sm-$A_B^*$ phase above a 1st order Sm-$A_B^*$--Sm-$C^*$ transition at $T_{AB}$. (i) is above the critical temperature $T_E>T_{AB}$ and (ii) is at $T_{AB}<T<T_E$ (see Fig.~\ref{PhaseDiagram} for the corresponding region in the phase diagram.) Curve (iii) is in the Sm-$C^*$ phase, below a 2nd or 1st order transition. 
\vspace{-25 pt}
}
\label{EE}
\end{figure}

The electroclinic effect (EE) in the Sm-$A^*_B$ phase is similar to that in a Sm-$A^*$ phase. Note that the EE in any Sm-$A^*$ phase will lead to {\it both} non-zero biaxiality and tilt. A signature of the 2nd order Sm-$A^*$--Sm-$A^*_B$ transition will be a discontinuity of $\frac{d \chi_0}{d T}$ but not a divergence of $\chi_0(T)$, where $\chi_0=\frac{d \theta}{d E}_{_{E=0}}$ is the zero-field susceptibility. The rapid increase in $\chi_0$ upon entry to the Sm-$A^*_B$ phase, shown in Fig.~\ref{EE}, corresponds to an enhanced EE. In fact, $\frac{d \chi_0}{d T}$ will diverge as $T\rightarrow T_{AB-}$. This behavior at the Sm-$A^*$--Sm-$A^*_B$ transition is in contrast to that at the 2nd order Sm-$A^*$--Sm-$C^*$ (or Sm-$A^*$--Sm-$C_\alpha^*$) transition where $\chi_0(T)$ diverges. Instead $\chi_0(T)$ diverges at the 2nd order Sm-$A^*_B$--Sm-$C^*$ transition. Thus, measuring $\chi_0(T)$ at the transition from the Sm-$A^*$ phase could distinguish the Sm-$A^*_B$ and Sm-$C^*_\alpha$ phases.

If the Sm-$A_B^*$--Sm-$C^*$ transition is 1st order the divergence of $\chi_0(T)$ will be cut off. However, the EE will be dramatic above the transition temperature ($T_{BC}$) and akin to that of a Sm-$A^*$ phase near a 1st order Sm-$A^*$--Sm-$C^*$ transition \cite{BH, Saunders2}. As shown in Fig.~\ref{EE}, there is a superlinear growth of $\theta(E)$ and, below a critical temperature $T_E>T_{BC}$, discontinuities and hysteresis in $\theta(E)$ are expected (without switching the sign of $E$). One expects two/four (without/with hysteresis) associated polarization current peaks instead of one/two peaks for a surface stabilized Sm-$C^*$ phase. This unusually strong EE has also been observed \cite{Laux, Bourny} above the Sm-$C^*_\alpha$--Sm-$C^*$ transition and is generally attributed to a competition between ferro- and antiferroelectricity in the Sm-$C^*_\alpha$ phase. If a Sm-$C^*_\alpha$ phase was to be misidentified as a Sm-$A^*_B$ phase then such EE behavior could be due instead to the proximity of the 1st order Sm-$A_B^*$--Sm-$C^*$ transition. 

We will now briefly describe our theory. First we map out the phase diagram for the non-chiral Sm-$A$, Sm-$A_B$ and Sm-$C$ phases. The corresponding phase diagram for a chiral system will differ quantitatively (e.g., the exact location of the phase boundaries) but not qualitatively, i.e., the diagram's topology, the possible phase sequences, the order (1st or 2nd) of the transitions will remain the same. Thus, for the sake of clarity, the effects of chirality will be only considered when analyzing the manifestly chiral features, i.e. the helical superstructures and EE. 

The Sm-$A$, Sm-$A_B$, Sm-$C$ phases can be distinguished by their second rank tensor orientational order parameter $\cal Q$, which we express as a sum of uniaxial and biaxial parts: $Q_{ij}= \sqrt{\frac{3}{2}}S \left[ \cos(\alpha)U_{ij} + \frac{\sin(\alpha)}{\sqrt{3}}B_{ij}\right] $, where $U_{ij}=n_{i} n_{j}-\frac{1}{3}\delta_{ij}$ is the uniaxial part and $B_{ij}= e_{1i} e_{1j}-e_{2i} e_{2j}$ is the biaxial part. Taking the smectic layer normal to point along $\bf \hat z$, the eigenvectors are: ${\bf \hat e_1}=-\sin\phi(z) {\bf \hat x} + \cos\phi(z) {\bf \hat y}$, ${\bf \hat e_2}=\cos\theta\left(\cos\phi(z) {\bf \hat x} + \sin\phi(z) {\bf \hat y}\right)-\sin\theta {\bf \hat z}$ and ${\bf \hat n}=\sin\theta\left(\cos\phi(z) {\bf \hat x} + \sin\phi(z) {\bf \hat y}\right)+\cos\theta {\bf \hat z}$. They and the angles $\phi$ and $\theta$ are shown in Fig.~\ref{Schematic}. The parameter $\alpha$ corresponds to the degree of biaxiality. The overall orientational order is $S=\sqrt{ Tr({\cal Q}^2)}>0$. The Sm-$A$ phase is untilted ($\theta = 0$) and uniaxial ($\alpha=0$). The Sm-$A_B$ phase is untilted ($\theta = 0$) and biaxial ($\alpha\neq0$) while the Sm-$C$ phase is tilted ($\theta \neq 0$) and biaxial ($\alpha\neq 0$). In the Sm-$A_B^*$ or Sm-$C^*$ phases, a helical superstructure corresponds to $\phi(z)=2 \pi z/p$ with $p$ the pitch. 

To analyze the transitions between the three phases we use a mean field Landau free energy density which, to lowest order in $\alpha$ and $\theta$, is:
\begin{eqnarray}
f&=& \frac{r_\theta }{2} \theta^2 + \frac{u}{4} \theta^4 +\frac{ \theta^6}{6}  + \frac{r_\alpha}{2}  \alpha^2 + \frac{\alpha^4}{4}  -\gamma \alpha \theta^2\;.
\label{f}
\end{eqnarray}
$r_\theta(T)$ and $r_\alpha(T)$ are monotonically increasing functions of $T$, e.g.,  $r_\alpha(T)=a_\alpha(T-T_\alpha)$ and $r_\theta(T)=a_\theta(T-T_\theta)$, where $a_\alpha,a_\theta>0$, and $T_\alpha$, $T_\theta$ are the temperatures below which, for zero coupling ($\gamma=0$), $\alpha$ and $\theta$ each become nonzero. The coupling term's structure, linear in $\alpha$ and quadratic in $\theta$, is important. It reflects the fact that if the system has tilt order, then by symmetry it must also have biaxial order, but not vice versa. Both $u, \gamma>0$ but the coupling term will effectively reduce the $\theta^4$ coefficient, even making it negative. Thus, the $\theta^6$ term is required to stabilize the system. The simple form of the $\theta^6$ and $\alpha^4$ coefficients is achievable  by rescaling $\theta$ and $\alpha$. We note that the above $f$ can be obtained by directly expanding in powers of $Q_{ij}$ and a smectic layering order parameter, an approach which was taken in \cite{Saunders2}. However for the sake of brevity we do not take this approach here. 

The phase diagram in $r_\alpha$--$r_\theta$ space, shown in Fig.~\ref{PhaseDiagram}, is obtained by minimizing $f$ with respect to $\alpha$ and $\theta$. There are two tricritical points (TCPs), at each of which 1st and 2nd order phase boundaries (for the Sm-$A$--Sm-$C$ and Sm-$A_B$--Sm-$C$ transitions) meet, as well as a critical end point (CEP) where the continuous Sm-$A$--Sm-$A_B$, 1st order Sm-$A$--Sm-$C$ and Sm-$A_B$--Sm-$C$ boundaries meet. Reducing $T$ corresponds to moving from upper right to lower left. There are 4 qualitatively different paths. Paths (1) and (2) do not involve a Sm-$A_B$ phase and exhibit 2nd and 1st order Sm-$A$--Sm-$C$ phase transitions, respectively. The Sm-$A_B$ phase appears along paths (3) and (4), each with a continuous  Sm-$A$--Sm-$A_B$ transition. The Sm-$A_B$--Sm-$C$ transition is 1st and 2nd order for paths (3) and (4), respectively.

We analyze the Sm-$A_B^*$ and Sm-$C^*$ helical superstructures by adding to $f$ the term $f_{chiral} =-h\epsilon_{ijk}Q_{jl}\partial_i Q_{kl}$, where  $\epsilon_{ijk}$ is the Levi-Cevita symbol. $h$ depends on the enantiomeric excess and is zero in a racemic system. This term, which favors a chiral distortion, must be stabilized by the elastic terms: $f_{elastic} =\frac{k_t}{4}\partial_iQ_{jk}\partial_i Q_{jk} +\frac{k_b-k_t}{2}\partial_iQ_{ij}\partial_k Q_{kj}$, where $k_t$ and $k_b$ are proportional to the the twist and bend elastic modulii, i.e., $K_{t/b}=\frac{3}{2}k_{t/b} S^2$. In the Sm-$A_B^*$ or the Sm-$C^*$ phases $f_{chiral}+f_{elastic}$ is minimized by $\phi(z)=2 \pi z/p$ \cite{fluctuation footnote} with pitch $p(T)$: 
\vspace{-5 pt}
%
%
%
\begin{eqnarray}
p(T)=2\pi\frac{k_t}{h}\left(\frac{1+\kappa x^2(T)}{1+x^2(T)}\right) \;,
\label{pitch}
\end{eqnarray}
with $\kappa=k_b/k_t$ and $x(T)=\theta(T)/\alpha(T)$. In the Sm-$A_B^*$ phase ($\theta=0$), $p=p_B=2\pi k_t/h$. Setting $\alpha=0$, one gets the usual uniaxial Sm-$C^*$ pitch $p_C=2\pi k_b/h$. The ratio $p_C/p_B=K_b/K_t$ is the ratio of the energies of each helical distortion. In the Sm-$A^*_B$ phase the distortion is a twist of the biaxial director ${\bf \hat e_1}$. In the uniaxial Sm-$C^*$ phase the higher energy distortion is a bend of the uniaxial director ${\bf \hat n}$. We note that the pitch lengths are equal in a one constant ($k_t=k_b$) approximation. Generally the Sm-$C^*$ pitch lies between $p_B$ and $p_C$. $p(T)$ for the sequence Sm-$A^*$--Sm-$A_B^*$--Sm-$C^*$, is summarized in Fig.~\ref{PhaseDiagram}. Upon entry to the Sm-$A_B^*$ phase $p(T)=p_B$ and remains constant. Entering the Sm-$C^*$ phase along path (4) (via a continuous Sm-$A_B^*$--Sm-$C^*$ transition), $p(T)$ will increase continuously towards $p_C$ as the $\theta^2(T)/\alpha^2(T)$ terms in Eq.~(\ref{pitch}) grow. Path (3) involves a 1st order Sm-$A_B^*$--Sm-$C^*$ transition where both $\alpha$ and $\theta$ jump. The most dramatic behavior occurs in the limiting case $u,\gamma\ll1$, where $\theta^2 \gg \alpha^2$ upon entry to the Sm-$C^*$ phase. Here $p(T)$ jumps, by a factor $\approx K_b/K_t$, up to $p\approx p_C$. 

The reduction in birefringence $\Delta n$ can be obtained by position averaging $Q_{ij}$ over the helical pitch. Using $\Delta n \propto \sqrt{ Tr({\cal Q}^2)}$ we find that the fractional reduction in $\Delta n$, is: 
\vspace{-10 pt}
\begin{eqnarray}
\Delta_{\Delta n}(T)=\frac{\alpha(T)^2}{2} + \frac{3\theta(T)^2}{2}\;.
\label{birefringence general}
\end{eqnarray}
Thus, as well as a decrease in $\Delta n$ going from the Sm-$A^*$ to the Sm-$A_B^*$ phase, one will observe a decrease going from Sm-$A_B^*$ to the Sm-$C^*$ phase when $\theta(T)$ becomes non-zero. For a 1st order transition one will observe a jump in $\Delta_{\Delta n}(T)$. The material MHPOBC shows the latter behavior \cite{Skarabot}. Whereas Ref.\cite{Skarabot} attributes this to some sort of structural change at the Sm-$C_\alpha^*$--Sm-$C^*$ transition, it could more simply attributed to the development of tilt order (in addition to biaxial order) at a 1st order Sm-$A_B^*$--Sm-$C^*$ transition. 

Our analysis of the electroclinic effect (EE) is preliminary in that we do not consider the role played in the EE by a possible helical superstructure. Keeping in mind that the layer normal points along $\bf \hat z$, we add the following term to $f$: $f_{EE} =e'\epsilon_{zjk}E_jQ_{zk}\approx -eE_\perp\theta$ where ${\bf E_\perp} \perp \bf {\hat z}$, $e=\sqrt{\frac{3}{2}}Se'$. For a racemic mixture $e' = 0$. The $\approx$ means we do not consider effect of ${\bf E}$ on $\alpha$ \cite{EE footnote}. Two features of the EE are as follows. Firstly, in the Sm-$A^*$ and Sm-$A_B^*$ phases, near the 1st order transitions to the Sm-$C^*$ phase (i.e. $T_{AC/BC}<T<T_E$), a discontinuous and hysteretic $\theta(E)$ (see Fig.~\ref{EE}) is observed. Figure~\ref{PhaseDiagram} shows the locus in $r_\alpha$-$r_\theta$ space corresponding to the critical temperature $T_E$. The second feature is the behavior of $\chi_0(T)$ in the phase sequence Sm-$A^*$--Sm-$A_B^*$--Sm-$C^*$. Outside the Sm-$C^*$ phase, $\chi_0^{-1}$, (see Fig.~\ref{EE}) is
\vspace{-10 pt}
\begin{eqnarray}
\chi^{-1}_0=
\begin{cases}
 a_\theta\left(T-T_\theta\right)& \text{$T>T_{AB}$} \\
a_\theta\left(T-T_\theta\right)-2\gamma a_\alpha^\frac{1}{2}\left(T_{AB}-T\right)^\frac{1}{2}& \text{$T_{BC}<T<T_{AB}$} \\
\end{cases}&\;.
\vspace{-50 pt}
\label{chi}
\end{eqnarray}
$T_{BC}>T_\theta$ is the continuous Sm-$A_B^*$--Sm-$C^*$ transition temperature, given by $\chi^{-1}_0(T_{BC})=0$. If the Sm-$A_B^*$--Sm-$C^*$ transition is 1st order then the growth of $\chi_0^{-1}(T)$ is cutoff at the transition ($T_{BC1st}>T_{BC}$). For a sequence Sm-$A^*$--Sm-$C^*$, $\chi^{-1}_0=a_\theta\left(T-T_\theta\right)$ in the Sm-$A^*$ phase. 

In summary, we have presented the first analysis of the Sm-$A_B^*$ phase, which has a helically precessing biaxial director. The helical pitch will be significantly shorter than that of the Sm-$C^*$ phase. A decrease in the birefringence and a strengthening of the EE will be observed below the Sm-$A^*$--Sm-$A_B^*$ transition. For systems with a 1st order Sm-$A_B^*$--Sm-$C^*$ phase transition, an unusually strong EE (with switching and hysteresis) will be observed in the Sm-$A_B^*$ phase. The above features are shared by the Sm-$A^*_B$ and Sm-$C^*_\alpha$ phases and we propose that it is possible that the Sm-$A^*_B$ phase could be misidentified as the Sm-$C^*_\alpha$ phase. We have discussed ways to distinguish the two phases. This work was sponsored by the National Science Foundation under Grant No. DMR-1005834.


\begin{thebibliography}{99}
\bibitem{Osipov Biaxiality} The nonchiral Sm-$A_B$ phase has recently been studied using molecular field theory \cite{Teixeira}. Lacking chirality,  it exhibits neither a helical superstructure, nor the EE considered here. Also a Sm-$A_B$--Sm-$C$ transition was not considered.
\bibitem{critical point} Of course, one can go continuously between two symmetrically identical phases by going around a critical point. However, this is not a continuous phase transition, which occurs at a distinct temperature and exhibits well defined thermodynamic features. Thus, a continuous Sm-$C^*_\alpha$--Sm-$C^*$ phase transition would have to occur right at a critical point. Indeed, in one case \cite{Huang} it was suggested that such a continuous transition may be occurring at a critical point but it is seems unlikely that this is the case with all continuous Sm-$C^*_\alpha$--Sm-$C^*$ transitions.
\bibitem{fluctuation footnote} We do not consider spatial variations in $\theta$ or $\alpha$ but will do so when considering fluctuation effects on the transitions, an analysis which we will present in a future publication.
\bibitem{EE footnote} For small field $E$, $\theta$ and $\alpha$ scale like $E$ and $E^2\ll E$ respectively. Thus, we neglect the direct effect of $E$ on $\alpha$. However, $\alpha$ will be induced due its coupling to $\theta$.
\end{thebibliography}
\end{document}